\documentclass[mathleft
]{an}
\usepackage{graphicx}

\usepackage{array,longtable}

\usepackage{times}
\begin{document}

\Pagespan{1}{}
\Yearpublication{2021}%
\Yearsubmission{2020}%
\Month{3}%
\Volume{xxx}%
\Issue{xx}%

\title{An investigation of open cluster Melotte 72 using Gaia DR2}

\author{Yasser Hendy\thanks{Corresponding author:
    \email{y.h.m.hendy@gmail.com}}, Ashraf Tadross (altadross@gmail.com)}


\titlerunning{Open cluster Melotte 72}
\authorrunning{Hendy \& Tadross}
\institute{National Research Institute of Astronomy and Geophysics, 11421 - Helwan, Cairo, Egypt}

\abstract {\textbf{Abstract}. The estimation of the main parameters of star clusters is significant in astrophysical studies. The most important aspect of using the Gaia DR2 survey lies in the positions, parallax, and proper motions of cluster stars with homogeneous photometry
that make the membership probability determine with high accuracy. In this respect, depending on Gaia DR2 database, an analysis of the open star cluster Melotte 72 is taking place here. It is located at a distance of 2345$\pm$108 pc with an age of 1.0$\pm$0.5 Gyr. In studying the radial density profile, the radius is found to be 5.0$\pm$0.15 arcmin. The reddening, the luminosity and mass functions, the total mass of the cluster, and the galactic geometrical distances ($X_\odot$, $Y_\odot$, $Z_\odot$), and the distance from the galactic center ($R_g$) have been estimated as well. Our study has shown a dynamical relaxation behavior of Melotte 72.
}


\received{3 October 2020} \accepted{xx xxx 2020} \publonline{later}

\maketitle

\section{Introduction}
Studying open star clusters (OCs) are playing an important role in understanding the galactic structure and evolution of the Milky Way Galaxy (cf. Gilmore et al. 2012; Moraux 2016). Our knowledge of OC's parameters, e.g., diameters, distances, ages, and color excesses are needed to obtain reliable conclusions. The information about OCs is compiled in large catalogs and databases, e.g., Dias et al. (2002, 2014) and Webda
(https://webda.physics.muni.cz/navigation) (cf. Roeser et al. 2010; Kharchenko et al. 2013; Dias et al. 2014; Sampedro et al. 2017). Kharchenko et al. (2013) extended the basic data for a large set of clusters derived in a uniform manner. Netopil et al. (2015) compared the clusters’ parameters in many catalogs and concluded that there are discrepancies in distances, color excesses, and ages.

Using Gaia DR2 survey, we believe that a large set of unknown clusters will be discovered in the Gaia era (cf. Castro-Girard et al. 2020 and Cantat-Gaudin et al. 2020). Gaia is a high-precision database for 1.7 billion sources in the astrometric parameter of coordinates, proper motions, and parallaxes ($\alpha$, $\delta$, $\mu_{\alpha}$ $cos\delta$, $\mu_{\delta}$, $\pi$). In addition, magnitudes in three photometric filters ($G$, $G_{BP}$, $G_{RP}$) for more than 1.3 billion sources (Gaia Collaboration 2018). Gaia Archive is available through the web page \\(http://www.cosmos.esa.int/gaia).

Melotte 72 is defined as a small, compressed cluster in constellation Monoceros, lies $\sim$ 1.3 degrees south-west of $\alpha$ Mon., as shown in Fig. 1. Hasegawa et al. (2008) studied the BVI photometric of 36 old open clusters using the 65-cm telescope at Gunma Astronomical Observatory. They evaluated the main parameters of Melotte 72 and found this object has the age of 1.6 Gyr, located at a distance of about 3175 pc and has a reddening of 0.10 mag. Two years later, Piatti et al. (2010) obtained the CCD UBVI photometry of 5 clusters included Melotte 72, they found that it has an intermediate-age of 600 Myr, located at a distance of 3000 pc and has reddening of 0.20 mag. According to Collinder (1931), it was thought that Melotte 72 is the same object as Collinder 467! Cantat-Gaudin et al. (2020) studied the distance, age, and interstellar reddening for about 2000 clusters identified with Gaia DR2 astrometry included Melotte 72. However, our results are in better agreement with the Cantat-Gaudin et al. (2020) work than with the other prior results.

Mass function study of OCs is a very important parameter in understanding the star formation processes (Kroupa 2001, Elmegreen 1999, Richtler 1994). Color magnitude diagram of the cluster is used to derive the luminosity function and mass function. Considerable work has been accomplished by many researchers (Phelps \& Janes 1993, Piskunov et al. 2004, Scalo 1986, Elmegreen 2000, Yadav \& Sagar 2002, 2004a). In this paper, one of our main aim is to determine the mass function of Melotte 72 and re-estimated its main astrophysical properties using the database of GAIA mission. We determined the membership probability (MP) of the cluster stars using the Automated Stellar Cluster Analysis (ASteCA) code (Perren et al. 2015, 2020).

This paper is organized as follows: Astrometric Information are presented in Section 2. The cluster center and diameter are depicted in Section 3. The photometry of the color magnitude diagram is performed in Section 4. Luminosity - mass functions and the dynamical status of the cluster are described in Section 5. Finally, the conclusions of our study are summarized in Section 6.

\begin{figure}\resizebox{\hsize}{!}
{\includegraphics[width=7cm,height=6cm]{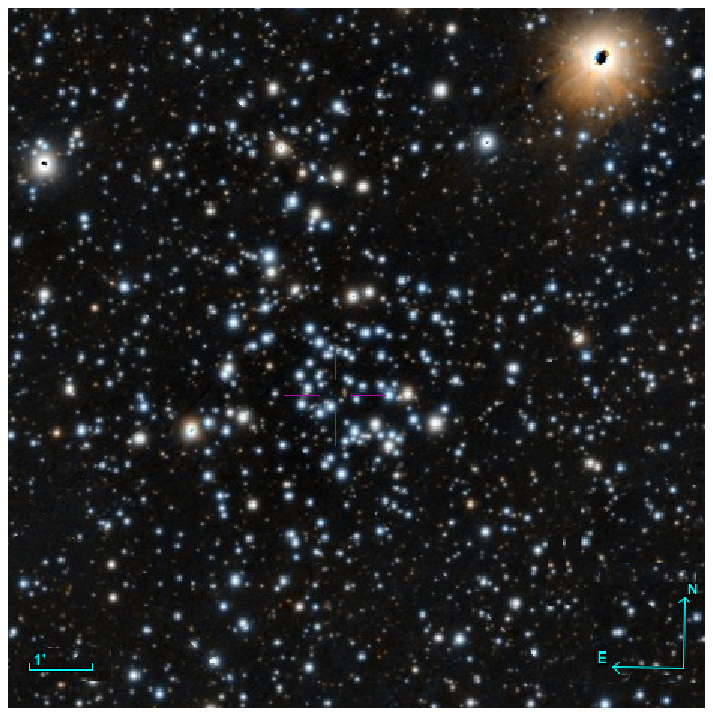}}
\caption{Image of Melotte 72 as taken from ALADIN. It lies in constellation Monoceros. A scale of 1 arcmin is shown on the lower left side of the figure. North is up, and East is to the left.}
\label{label1}
\end{figure}

\section{Astrometric Information}

To get the coordinates of Melotte 72, we used the Dias et al. (2002) catalog (V. 2016; https://wilton.unif ei.edu.br/ocdb). The source data were downloaded from Gaia DR2 database service, located at the object’s center with a radius of 20 arcmin to cover all the cluster’s region, and contain all the needed data. Using the known astronomical tool of (TOPCAT), www.star.bristol.ac.uk/mbt/topcat/, the vector point diagram (VPD) of the proper motion in right ascension and declination (pmRA, pmDE) is plotted as shown in Fig. 2.

It is evident in the VPD that the Melotte 72 are well separated from the field stars. The radius of the circle was determined by plotting the stellar density as a function of radial distance in the VPD, see Fig. 3. We put a cut-off where stellar density falls close to the field density which is found to be 0.68 mas/yr and shown by the vertical dashed arrow in Fig. 3. The over density region is selected as a subset called the co-moving stars, i.e., the stars selected for the current study. In Fig. 4, we confirmed that those stars are moving together at the same speed and direction in the sky compared to the field stars, where it is one of our membership criteria we applied, Tadross (2018). Besides, they form a clear specific track on the color magnitude diagram (CMD), as well. On the other hand, the parallax range of the cluster can be seen in the left-hand panel of Fig. 5. In addition, the relation between the magnitudes and parallax is plotted, as shown in the right-hand panel of Fig. 5. It shows that the distance range is narrow for the cluster members, while it is larger for the background field stars. On this respect, a star is counted as a cluster member when located within the cluster’s area (estimated from the radial density profile in section 3), having the same velocity in the sky compared to the background field stars, and its 3$\sigma$ parallax error lies within the mean parallax of the cluster. To enhance the cluster data, the stars of parallax errors $\leq$ 0.2 mas, and proper motion errors $\leq$ 0.4 mas/yr are taken into account. The mean proper motions in RA and DE, which are found to be $-4.15$$\pm$0.20 mas/yr and $3.68$$\pm$0.18 mas/yr, repectively as shown in Fig.6.

ASteCA employs a Bayesian field star decontamination algorithm to compare the photometric, parallax, and proper motion distribution of the stars in the cluster region with a similar distribution in the surrounding field region (Perren et al. 2015, 2020). We used the astrometric data of Gaia DR2 to obtain MP, see Fig. 7. We considered the stars of MP $>$ 70$\%$ are most probable members.

$$$$$$$$\begin{figure}\resizebox{\hsize}{!}
{\includegraphics[width=7cm,height=6cm]{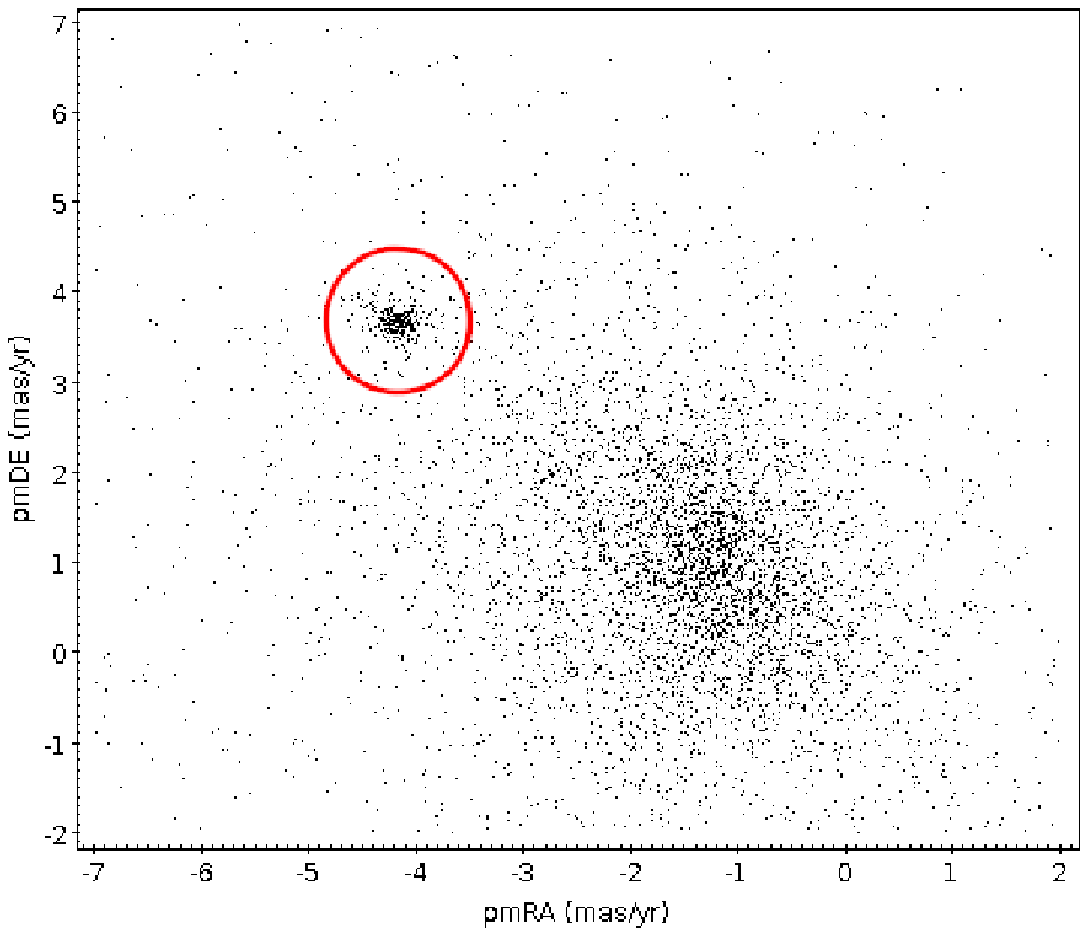}}
\caption{The vector point diagram VPD of Melotte 72 from which the stars in the highest concentrated area are selected for the current study.}
\label{label1}
\end{figure}

\begin{figure}\resizebox{\hsize}{!}
{\includegraphics[width=8.5cm,height=6cm]{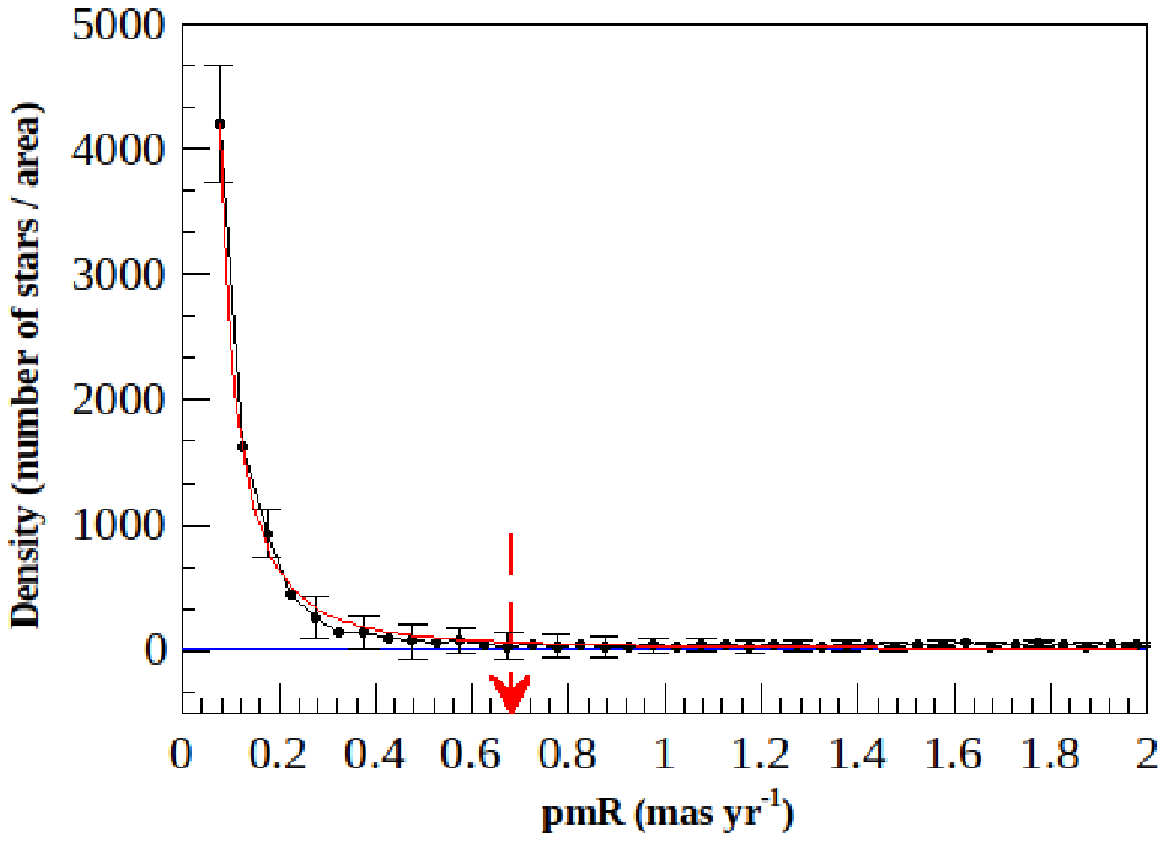}}
\caption{The radial distribution of stellar number density in the proper motion plane on the VPD. The solid horizontal line represents the field density and the vertical dashed arrow indicates the cut-off radius used to find the cluster members.}
\label{label1}
\end{figure}

\begin{figure}\resizebox{\hsize}{!}
{\includegraphics[width=8cm,height=6cm]{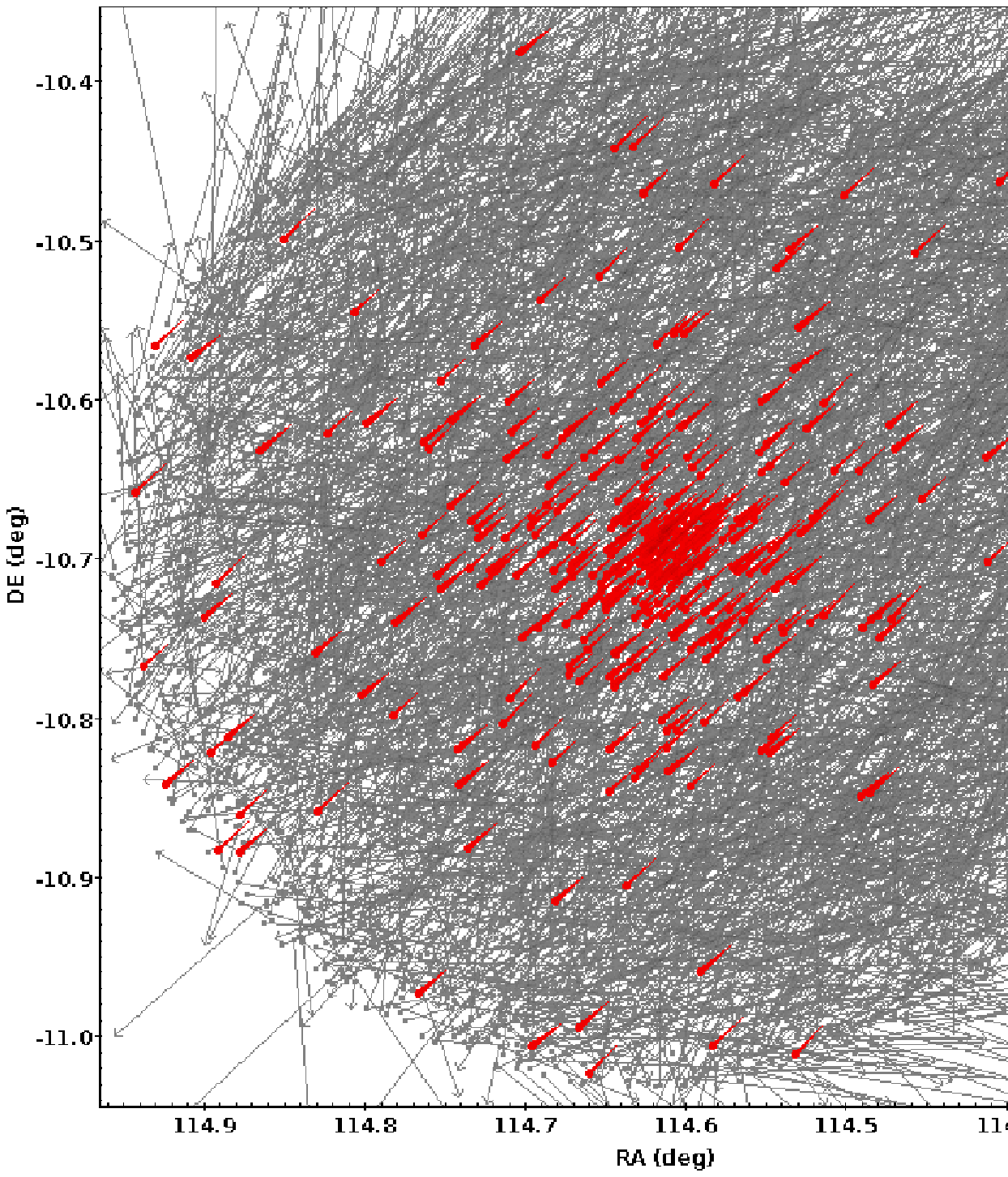}}
\caption{The co-moving stars of Melotte 72, i.e., the subset selected from VPD. It declares the stars that move together with the same  velocity in the sky.}
\label{label1}
\end{figure}

\begin{figure}\resizebox{\hsize}{!}
{\includegraphics[width=8cm]{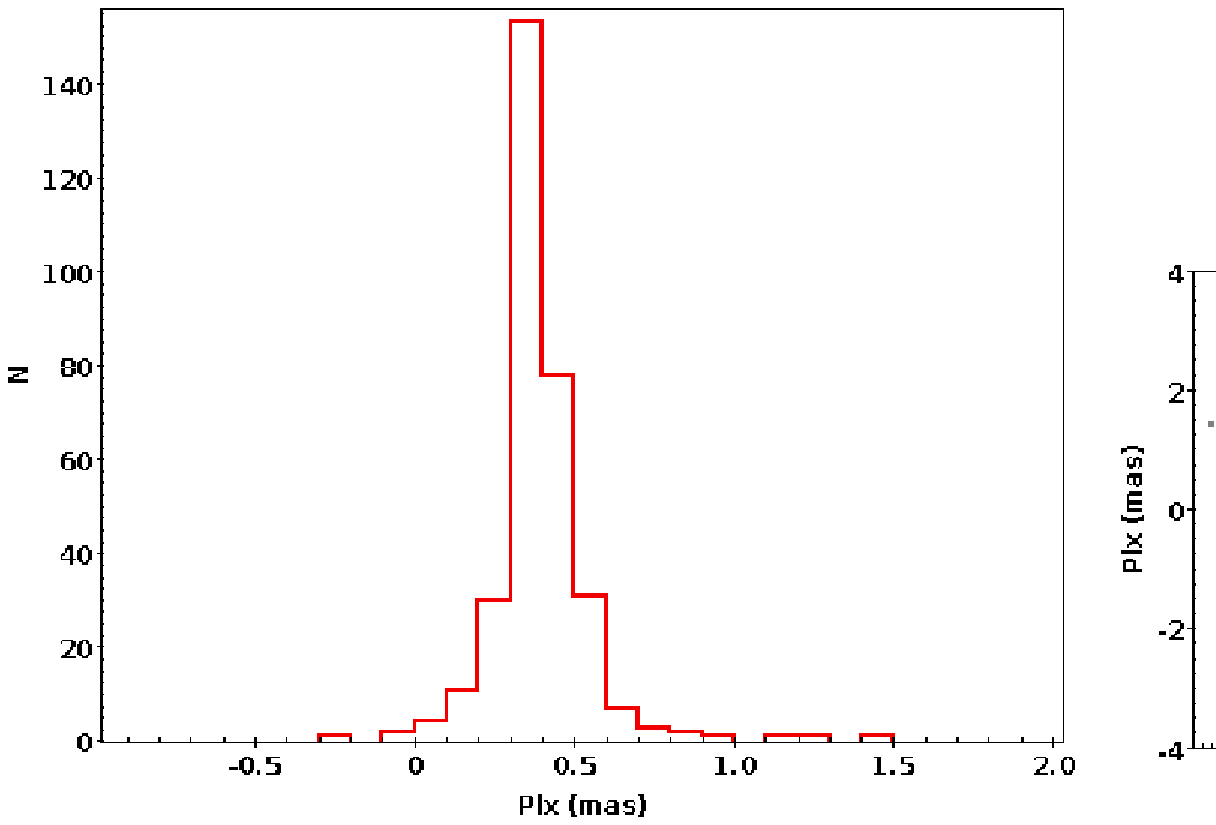}}
\caption{The left-hand panel shows the parallax range of Melotte 72. The right-hand panel presents the relation between the magnitude and parallax of Melotte 72. It shows that the distance range is narrow for the cluster members, while it is larger for the background field stars.}
\label{label1}
\end{figure}

\begin{figure}\resizebox{\hsize}{!}
{\includegraphics[width=8.5cm,height=6cm]{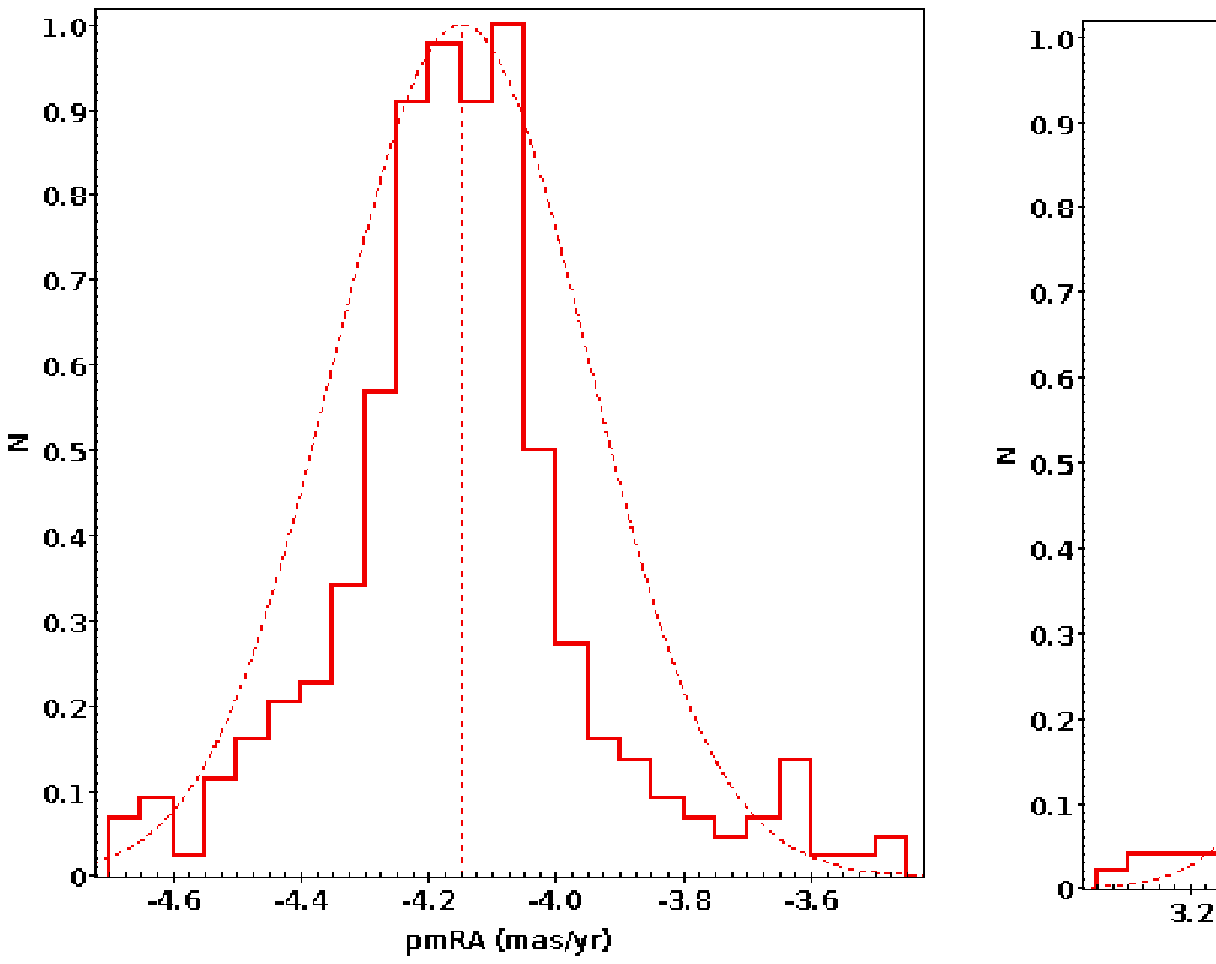}}
\caption{The mean proper motions in RA and DE of Melotte 72, which are -4.15$\pm$0.20 mas/yr and 3.68$\pm$0.18 mas/yr respectively.}  
\label{label1}
\end{figure}

\begin{figure}\resizebox{\hsize}{!}
{\includegraphics[]{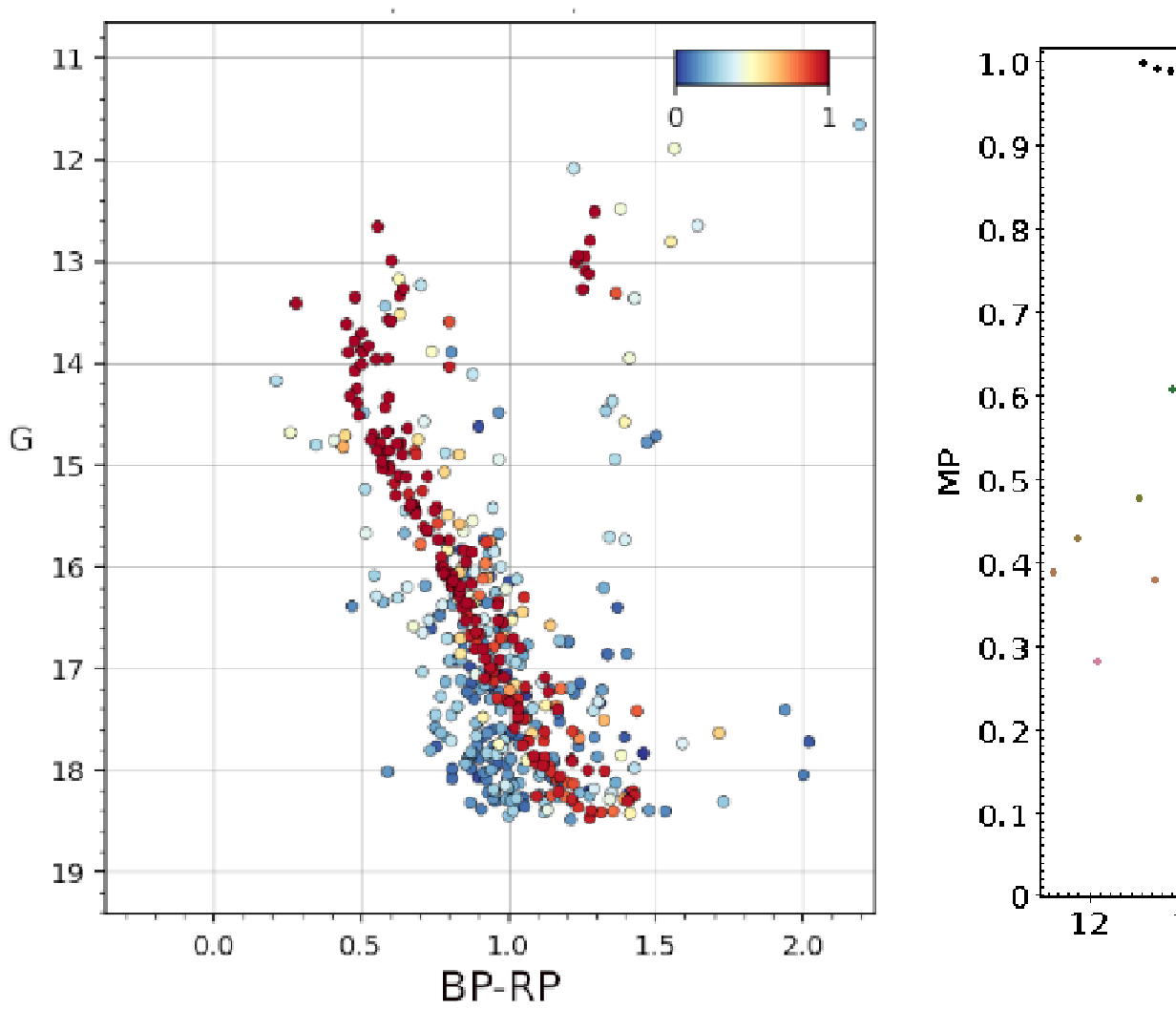}}
\caption{The left-hand panel shows the CMD with color density of membership probabilities. The right-hand panel presents G magnitude with membership probabilities from 0 to 1 using the ASteCA code.}
\label{label1}
\end{figure}

\section{Cluster’s structure (center and diameter) }

Usually, the cluster center is taken at the highest stellar density of the cluster’s region. Dividing the area under study into equal-sized bins in RA \& DE and counting the stars in each bin for both directions. A histogram of star counts is sketched, as shown in Fig. 8.
Fitting of Gaussian profiles are applied to both directions (RA \& DE). The obtained central coordinates of Melotte 72 are found to be 114.63$\pm$0.10 deg, and -10.7$\pm$0.09 deg in RA and DE, respectively, which agree well with Simbad and Dias (2002) Catalogs, and Cantat-Gaudin (2020), see Table 1.

To obtain the diameter of a cluster, the radial density profile (RDP) should be plotted for that cluster. The surface stellar density is derived by counting stars in concentric rings (shells) around the cluster’s center. Dividing the numbers of stars by their corresponding
shells’ areas, we can get the calculated density in each shell as illustrated before (cf. Tadross, 2005, 2009, 2018). As shown from the density distribution of Fig. 9, there is a peak near the cluster center and then decreases and flats after a certain point where the
cluster density dissolved into the field stars. The cluster radius, $R_{cluster}$ can be estimated at that value, which covers the whole cluster area and reaching sufficient constancy. King’s model (1966) is applied to the cluster according to the empirical relation:

\begin{equation}
f(R_{cluster}) = f_{bg} +\frac{f_{0}}{1+(R/R_{c})^2}\\
\end{equation}

where $f_{bg}$, $f_{0}$ and $R_{c}$ are background, central density, and core radius, respectively. The core radius is the distance at which the stellar density equals half the central density. By fitting the King model to the radial density profile, the estimated radius is found to be $R_{cluster}$ =5.0$\pm$0.15 arcmin, and the core radius $R_c$ =0.45$\pm$0.09 arcmin. On the other hand, the tidal radius of an open cluster is the distance from the cluster center at which the gravitational impact of the Galaxy equivalents to that of the cluster core. Calculating the overall mass of Melotte 72 (see Sec. 5), the tidal radius can be calculated using Jeffries (2001) equation, $R_t$ = 1.46 x (Mc)$^{1/3}$, where $R_t$ and $M_c$ are the tidal radius (in parsecs) and the overall mass of the cluster (in solar mass), respectively. It is found to be $R_t$ =9.7 pc. Also, the concentration parameter, which is an important parameter, shows us how the cluster is prominent in the sky. It can be calculated using the equation C=log ($R_{cluster}$ /$R_c$ ) of Peterson and King (1975). It is found to be C=1.3, which declares that Melotte 72 is a prominent cluster among the background field stars.

\begin{figure}\resizebox{\hsize}{!}
{\includegraphics[width=15 cm, height=10 cm]{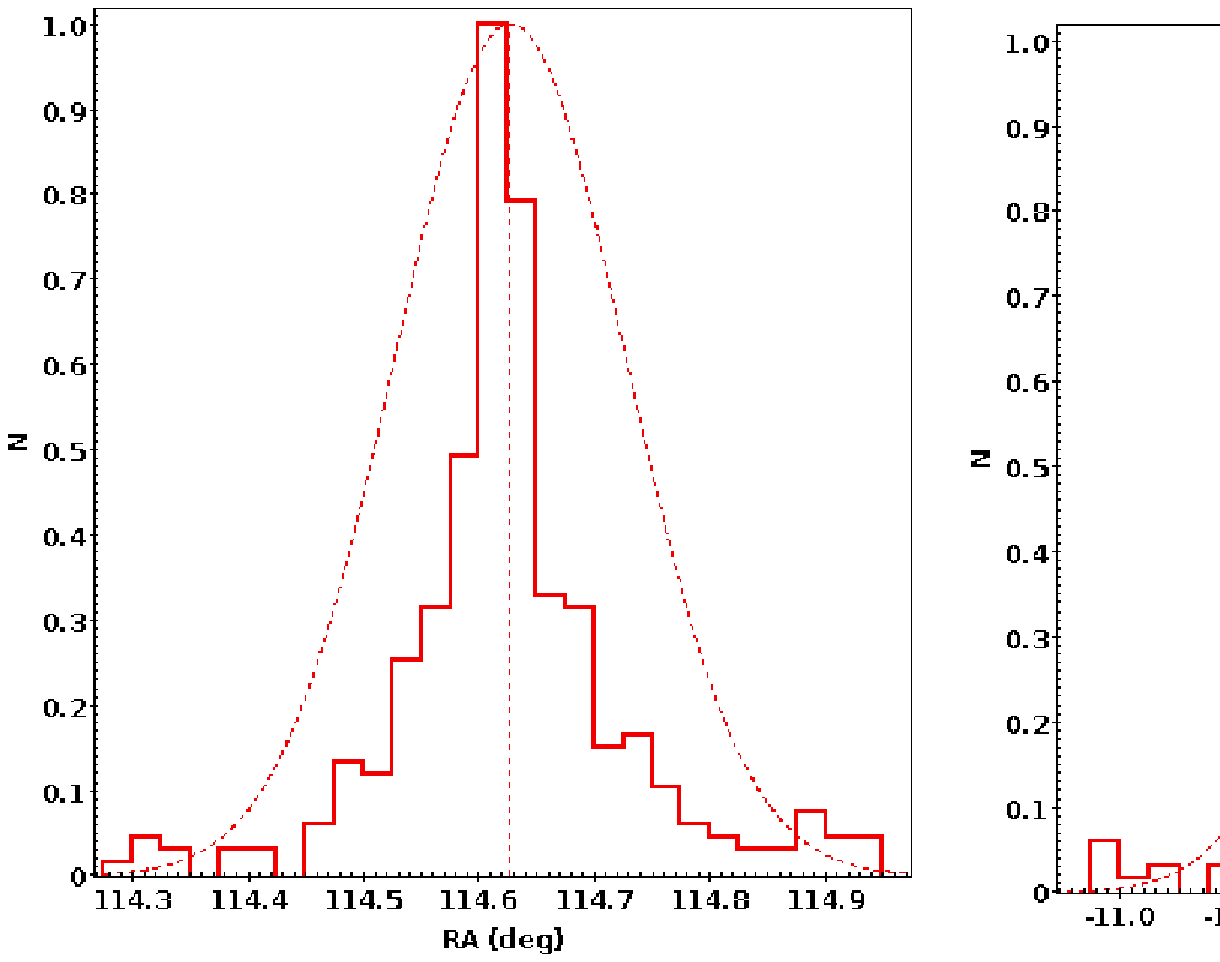}}
\caption{Cluster center of Melotte 72, the mean values are found to be 114.63$\pm$0.10 deg, and -10.7$\pm$0.09 deg in RA and DE, respectively.}
\label{label1}
\end{figure}

\begin{figure}\resizebox{\hsize}{!}
{\includegraphics[width=8.5cm,height=6cm]{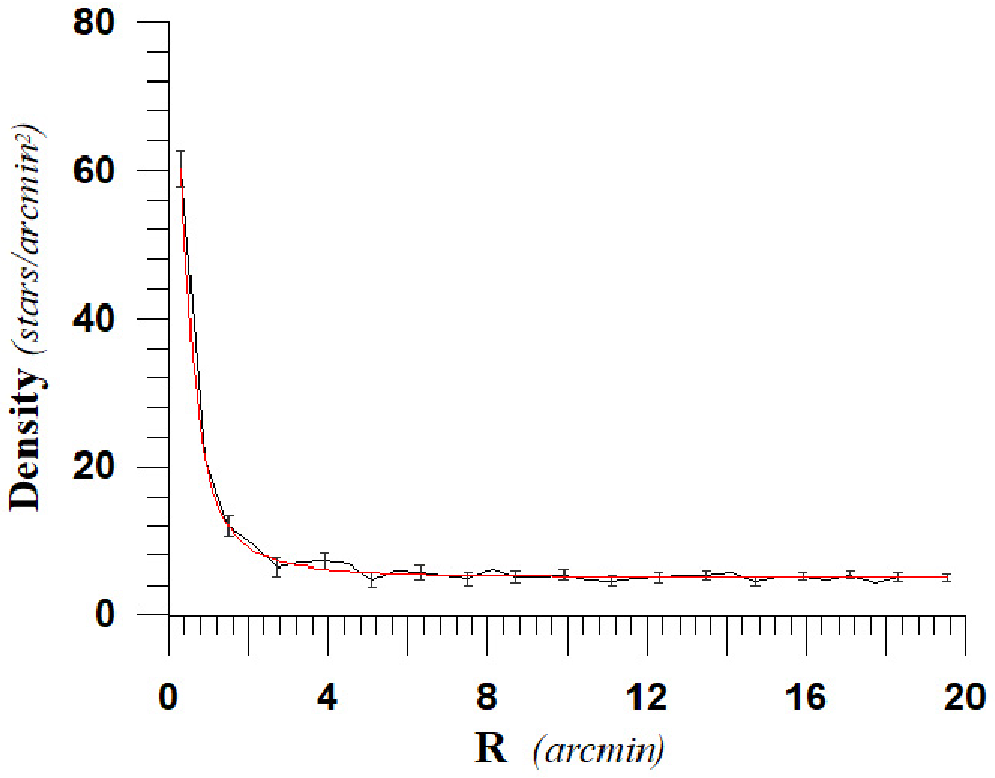}}
\caption{Surface density profile of Melotte 72. The error bars refer to the Poisson distribution, and the red line refers to the fitted profile of King (1966). The estimated radius was found to be 5.0$\pm$0.15 arcmin.}
\label{label1}
\end{figure}

\section{Color-Magnitude Diagram}

CMD is an essential tool for estimating the fundamental parameters of the candidate cluster. Age, distance modulus, color excess, and  metallicity can be estimated all at once by applying the theoretical stellar isochron to the observed CMD and getting the best fit. It is substantial to recognize that the cluster's region is contaminated by field stars.

For verifying our main results on ground-based observations, a matching file has been created between our membership stars and Piatti et al. (2010) ones, which are available online. The CMDs, (G$\sim$BP-RP) and (V$\sim$V-I) have been constructed, and we get the best fit of the zero-age main sequence to both CMDs, at the same age and the same distance modulus. The theoretical stellar isochrons were obtained from Marigo
et al. (2017). We used the recent theoretical isochron of nowadays solar metallicity of Z=0.0152 (Caffau et al. 2009, 2011 \& Steffi et al. 2018). According to the visual best of the two CMDs of Melotte 72, we got the age=1.0$\pm$0.5 Gyr, distance modulus, m-M=12.22$\pm$0.1 mag, and color excess, E(BP-RP)=0.18$\pm$0.05 mag, as shown in Fig. 10. On the other hand, the mean parallax of the members is found to be 0.38$\pm$0.09 mas, which yields a distance value agrees well with our fitting. Those members lie within the cluster diameter and located in the ranging parallax, within parallax’s errors $\leq$0.2 mas, proper motions’ errors in RA \& DE $\leq$0.4 mas/yr, and membership probability MP $>$ 70$\%$. Correspondingly, the Galactocentric Cartesian coordinates ($X_{\odot}$, $Y_{\odot}$, $Z_{\odot}$), and the distance from the galactic center ($R_g$) are estimated for Melotte 72 (cf. Tadross 2011) and listed in Table 1.


\begin{figure}
\begin{center}
\includegraphics[width=8cm,height=18cm]{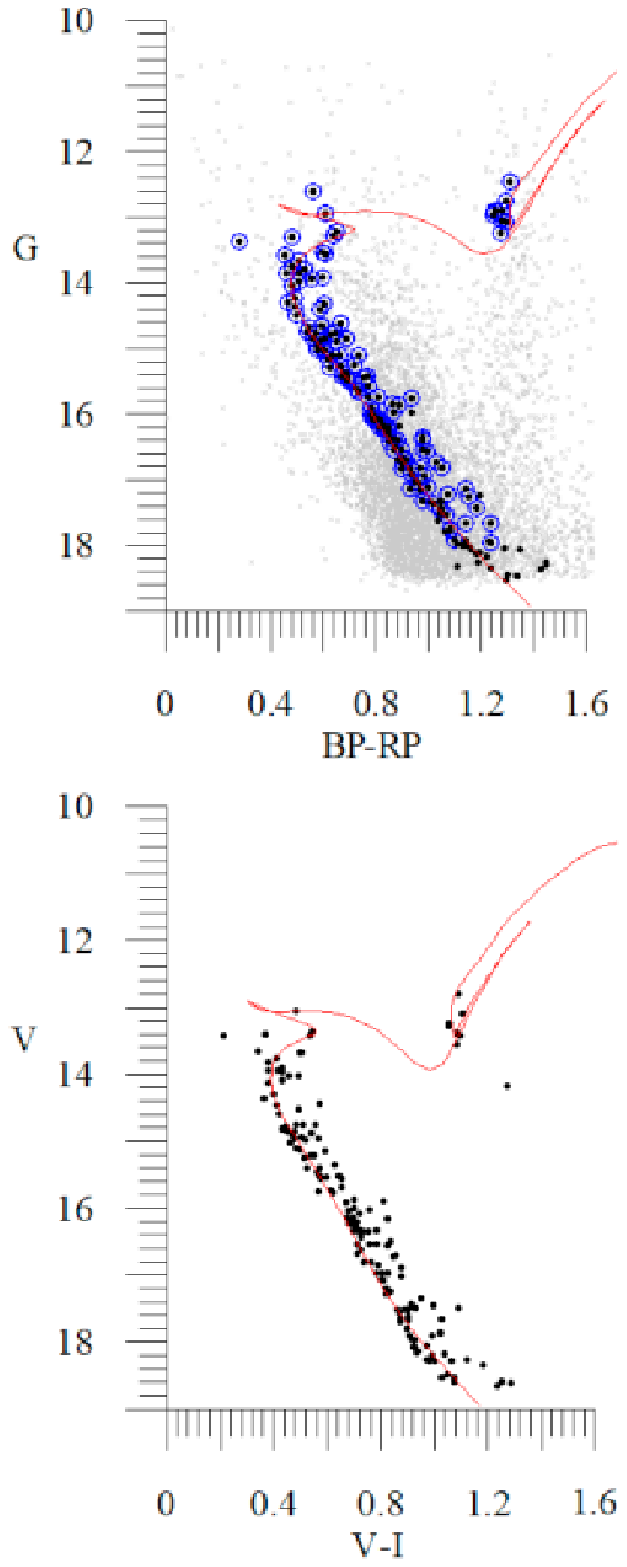}
\caption{The upper panel shows the CMD of our members that matched with Cantat-Gaudin et al. (2020) by 132 (blue open circles) of 168 (black dots) stars. The lower panel shows the V$\sim$(V-I) diagram of the ground-based observations of Piatti et al. (2010) that matched with ours by 159 of 168 stars. The theoretical isochron of solar metallicity Z=0.0152 aged 1.0 $\pm$0.5 Gyr of Marigo et al. (2017) has been applied to both diagrams. The visual best fit is taken at the same distance modulus of 12.22 $\pm$0.10 mag, and color excess of 0.18 $\pm$0.05 mag. Our members (168 stars in Table 2) lied within the cluster’s size, in the same range of the parallax; taken the parallax errors $\leq$ 0.2 mas; proper motion errors $\leq$ 0.4 mas, and the membership probability MP $>$ 70$\%$.}
\label{error_pm}
\end{center}
\end{figure}


\begin{table*}
\caption{The astrophysical parameters of Melotte 72.}
\begin{tabular}{lllllllll}
\hline\noalign{\smallskip}
\smallskip\smallskip  
Parameter & Present study & Cantat-Gaudin et al. (2020)  & Piatti et al. (2010)&  Hasegawa et al. (2008) \\
\hline\noalign{\smallskip}
$\alpha$(2000)~~~$deg$    &  114.63$\pm$0.10  & 114.618 &---                 &---     \\
$\delta$(2000)~~~$deg$    &  -10.70$\pm$0.09  & -10.698 &---                 &---     \\
$l$(2000)~~~$deg$         &  227.870          & 227.862 &---                 &---     \\
$b$(2000)~~~$deg$         &  5.378            & 5.369   &---                 &---     \\
Age~~~$Gyr$               &  1.0$\pm$0.5      & 0.97    &0.60                &1.6     \\
$Metallicity (z)$~~~      &  0.0152           & 0.0196  &---                 &0.008     \\
E(BP-RP)~~~$mag$          &  0.18$\pm$0.05    &---      &---                 &---       \\
E(B-V)~~~$mag$            &  0.14             &---      &0.20                &---        \\
E(V-I)~~~$mag$            &  0.18$\pm$0.05    &---      &0.25                &0.10        \\
(m-M)~~~$mag$             &  12.22$\pm$0.1    &12.14    &13.00               &12.51        \\
$A_v$~~~$mag$               &  0.43             &0.41     &---                 &---        \\
dist.~~~$pc$              &  2345             &2684     &3000                &3175       \\
Plx.~~~$mas$              &  0.38$\pm$0.09   &0.369$\pm$0.055      &---                &---        \\
PmRA~~~$mas/yr$           &  -4.15$\pm$0.20 &-4.159$\pm$0.110     &---                &---        \\
PmDE~~~$mas/yr$           &  3.68$\pm$0.18  &3.680$\pm$0.093      &---                &---        \\
PmR~~~$mas/yr$            &  0.68            &---     &---                &---        \\
Mem~~~$stars$             &  168               &177                  &---                &---     \\
$R_{cluster}$~~~$arcmin$            &  5.0$\pm$0.15     & $\sim$3.0           &---                &---     \\
$R_{c}$~~~$arcmin$        &  0.45$\pm$0.09    & ---                 &---                &---     \\
$C$~~~                    &  1.15             & ---                 &---                &---     \\
$R_g$~~~$kpc$             &  10.60            & 10.33               &---                &10.88     \\
$X_\odot$~~~$pc$          &  -2018.94         & -1793               &---                &---     \\
$Y_\odot$~~~$pc$          & -2232.07          & -1981               &---                &---     \\
$Z_\odot$~~~$pc$          &  283.36           & 251                 &---                &300     \\
$R_{t}$~~~$\textit{pc}$   &  9.7              & ---                 &---                &---     \\
IMF slope~~~              &  -2.20$\pm$0.10   & ---                 &---                &---     \\
Total lumin.~~~$mag$      &  -2.92            & ---                 &---                &---     \\
Total mass~~~$M_\odot$    &  223              & ---                 &---                &---     \\
Relax. time~~~$Myr$       &  $\sim$34         & ---                 &---                &---     \\
$\tau$~~~                 &  $\sim$29          & ---                 &---                &---     \\
\hline\noalign{\smallskip}
\end{tabular}
{\smallskip}
\end{table*}

\section{Luminosity \& mass functions, and dynamical state}
An open star cluster exemplifies hundreds of stars having the same age and chemical structures but different masses. The luminosity and mass functions (LF \& MF) of a cluster are counted on the amount of the membership of that cluster. The main problem for studying the LF and MF is to remove the field star contamination among the cluster’s members. Using the conditions of the RDP, parallax, proper motions, and membership
probability of, we can separate cluster members (168 stars) from the field’s ones, as listed in Table 2. Counting the stars within the main sequence envelope in terms of the absolute magnitude $M_G$ that can be derived from the distance modulus of the cluster. A histogram of LF for Melotte 72 can be drawn, as shown in the left-hand panel of Fig. 11. The luminosity and masses of the clusters’ members can be given from the isochron data of Marigo et al. (2017) at the age of the cluster. We infer that the massive stars are concentrated in the cluster center, while the peak lies at the fainter magnitude bin. The total luminosity of Melotte 72 is found to be -2.92 mag.

The LF and MF are inseparable according to the familiar relation called the mass-luminosity relation. Since we have not an empirical transformation for such a relationship here, we rely on the isochron data. The cluster members were divided into mass intervals of the absolute magnitude bins. The resulting MF of Melotte 72 is shown in the right-hand panel of Fig. 11. The linear fit represents the initial mass function IMF slope that obtained from the following equation:

\begin{figure}\resizebox{\hsize}{!}
{\includegraphics[]{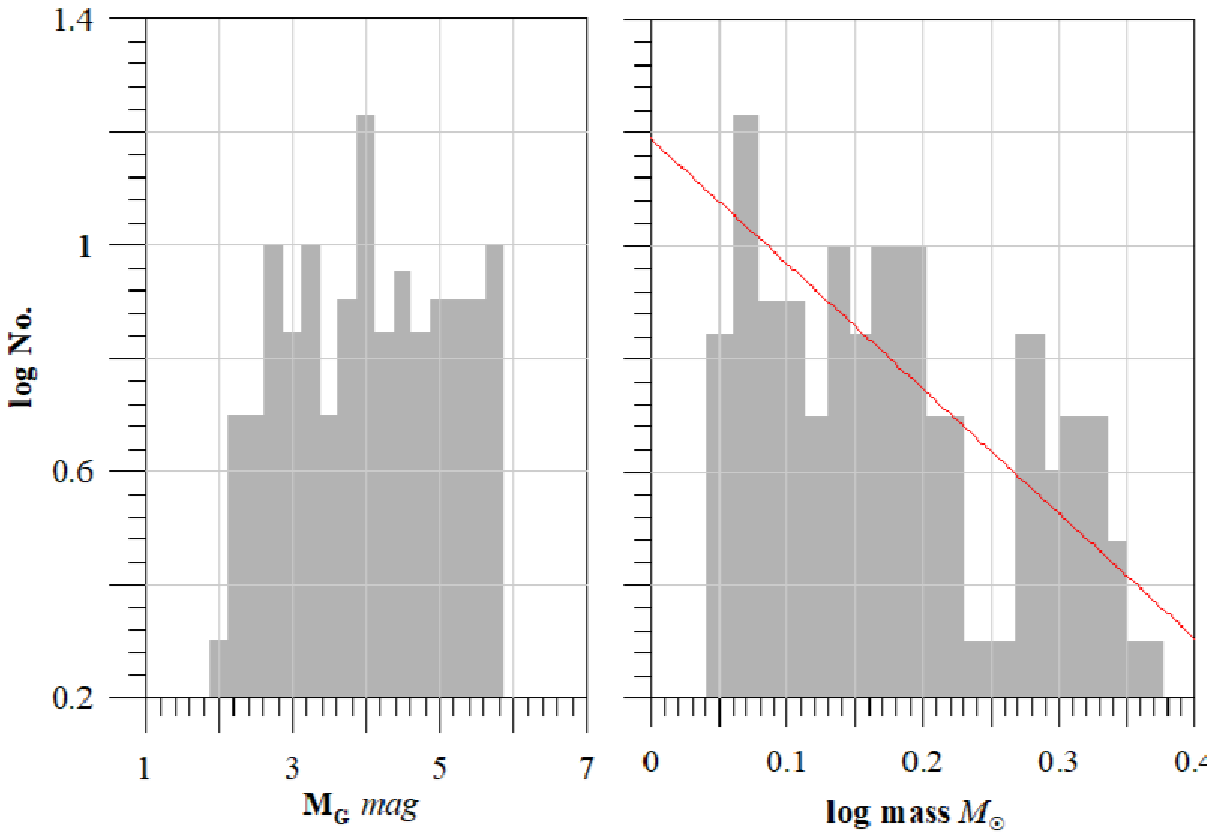}}
\caption{The left-hand panel represents the luminosity function distribution of Melotte 72. The right-hand panel represents the mass function distribution of the cluster, where the red linear fit shows the slope of the relation ($\alpha$ = -2.20$\pm$0.10). The total luminosity and masses were found to be -2.29 mag and 223 M$_{\odot}$, respectively.}
\label{label1}
\end{figure}

$\frac{dN}{dM} \propto {M^{-\alpha}}$ \quad \quad \quad \quad \quad \quad \quad \quad \quad \quad \quad \quad \quad \quad \quad \quad (2)




Where $\frac{dN}{dM}$  is the number of stars in the mass interval [M: (M+dM)] and $\alpha$ is the slope of the relation. It is found to be -2.20$\pm$0.10, which indicates that the estimated masses of the Melotte 72 lie in the range of Salpeter (1955). The total mass of the target cluster is calculated by integrating the masses of the cluster members (Sharma et al. 2006). It is found to be 223 $M_{\odot}$. Finally, we calculated the relaxation time of the cluster, which is the time scale when the cluster loses all traces of its initial conditions and represented by $T_R$. It is the characteristic time-scale for the cluster to reach the equivalent level of energy, it is given by Spitzer \& Hart (1971) as:

\begin{center}
	
$\large T_{R} = \frac{8.9 \times 10^{5} \sqrt{N} \times R_{h}^{1.5}}{\sqrt{\langle m \rangle} \times log (0.4 N)}$\\

\end{center}

where $N$ is the number of members, $R_{h}$ is the radius containing half of the cluster mass in parsecs (assuming that $R_{h}$ equals half of the cluster radius), $\langle m \rangle$  is the average mass of the cluster. Then, the relaxation time is found to be $T_R$ = 34.5 Myr for Melotte 72. Comparing the age of the cluster and its relaxation time indicates that the relaxation time is much smaller than the age of the cluster. From the dynamical evolution parameter $\tau$= Age/$T_R$, we conclude that Melotte 72 is dynamically relaxed.


\begin{center}
\begin{table*}[h]
\caption{The cluster members of Melotte 72.}
\label{TAB_COMP_ASTROPHOTO}
\centering
\begin{tabular}{cccccccccccc}
\hline
ID & RA    & DE    & Plx   & e\_{Plx} & pmRA     & e\_{pmRA} & pmDE     & e\_{pmDE} & Gmag  & BP-RP & MP \\
   & (deg) & (deg) & (mas) & (mas)   & (mas/yr) & (mas/yr) & (mas/yr) & (mas/yr) & (mag) & (mag) &    \\
\hline
  1 & 114.644 & -10.779 & 0.331 & 0.036 & -4.201 & 0.060 & 3.461 & 0.048 & 15.419 & 0.757 & 0.98\\
  2 & 114.614 & -10.774 & 0.421 & 0.081 & -4.081 & 0.131 & 3.468 & 0.098 & 16.812 & 0.887 & 0.98\\
  3 & 114.643 & -10.760 & 0.325 & 0.030 & -4.244 & 0.051 & 3.711 & 0.042 & 14.787 & 0.565 & 0.99\\
  4 & 114.647 & -10.774 & 0.309 & 0.033 & -4.348 & 0.057 & 3.606 & 0.044 & 15.182 & 0.614 & 0.96\\
  5 & 114.635 & -10.765 & 0.470 & 0.051 & -4.198 & 0.083 & 3.558 & 0.066 & 16.168 & 0.874 & 0.98\\
  6 & 114.631 & -10.768 & 0.574 & 0.164 & -3.808 & 0.267 & 3.595 & 0.204 & 18.213 & 1.172 & 0.94\\
  7 & 114.587 & -10.764 & 0.370 & 0.021 & -4.145 & 0.037 & 3.683 & 0.031 & 13.268 & 0.644 & 1.0\\
  8 & 114.590 & -10.753 & 0.648 & 0.159 & -3.602 & 0.250 & 3.460 & 0.193 & 18.068 & 1.174 & 0.87\\
  9 & 114.582 & -10.757 & 0.494 & 0.076 & -4.124 & 0.130 & 3.558 & 0.100 & 16.797 & 1.040 & 0.98\\
  10 & 114.597 & -10.757 & 0.383 & 0.053 & -4.262 & 0.084 & 3.667 & 0.066 & 16.086 & 0.812 & 0.99\\
  11 & 114.614 & -10.737 & 0.419 & 0.049 & -4.153 & 0.078 & 3.651 & 0.069 & 15.858 & 0.877 & 0.99\\
  12 & 114.608 & -10.750 & 0.288 & 0.095 & -4.313 & 0.155 & 3.763 & 0.121 & 17.023 & 0.952 & 0.95\\
  13 & 114.664 & -10.750 & 0.484 & 0.043 & -4.010 & 0.069 & 3.652 & 0.060 & 15.739 & 0.799 & 0.98\\
  14 & 114.661 & -10.757 & 0.441 & 0.060 & -4.132 & 0.096 & 3.744 & 0.082 & 16.380 & 0.965 & 0.99\\
  15 & 114.663 & -10.729 & 0.040 & 0.159 & -4.692 & 0.248 & 3.933 & 0.205 & 18.125 & 1.203 & 0.84\\
  16 & 114.687 & -10.737 & 0.384 & 0.086 & -3.981 & 0.135 & 3.671 & 0.118 & 17.091 & 0.987 & 0.98\\
  17 & 114.681 & -10.719 & 0.300 & 0.059 & -4.279 & 0.096 & 3.493 & 0.084 & 16.371 & 0.871 & 0.97\\
  18 & 114.675 & -10.741 & 0.215 & 0.089 & -4.097 & 0.144 & 3.710 & 0.121 & 17.230 & 1.136 & 0.96\\
  19 & 114.666 & -10.736 & 0.493 & 0.052 & -4.002 & 0.087 & 3.468 & 0.076 & 16.154 & 0.813 & 0.96\\
  20 & 114.627 & -10.728 & 0.367 & 0.052 & -4.124 & 0.083 & 3.758 & 0.068 & 16.187 & 0.837 & 0.99\\
  21 & 114.632 & -10.727 & 0.463 & 0.037 & -4.061 & 0.057 & 3.705 & 0.044 & 15.109 & 0.629 & 0.99\\
  22 & 114.625 & -10.741 & 0.410 & 0.042 & -4.118 & 0.074 & 3.702 & 0.059 & 15.733 & 0.763 & 0.99\\
  23 & 114.624 & -10.733 & 0.356 & 0.034 & -4.099 & 0.054 & 3.620 & 0.043 & 15.119 & 0.655 & 0.99\\
  24 & 114.617 & -10.729 & 0.590 & 0.112 & -4.112 & 0.183 & 3.852 & 0.151 & 17.298 & 1.034 & 0.92\\
  25 & 114.650 & -10.729 & 0.305 & 0.036 & -4.039 & 0.059 & 3.470 & 0.051 & 15.385 & 0.678 & 0.98\\
  26 & 114.632 & -10.736 & 0.377 & 0.027 & -4.215 & 0.046 & 3.646 & 0.037 & 14.643 & 0.659 & 0.99\\
  27 & 114.649 & -10.731 & 0.400 & 0.047 & -4.242 & 0.078 & 3.693 & 0.061 & 16.014 & 0.781 & 0.99\\
  28 & 114.644 & -10.720 & 0.278 & 0.050 & -4.245 & 0.082 & 3.654 & 0.066 & 16.083 & 0.784 & 0.97\\
  29 & 114.653 & -10.715 & 0.406 & 0.041 & -4.215 & 0.066 & 3.694 & 0.057 & 15.643 & 0.724 & 0.99\\
  30 & 114.671 & -10.712 & 0.385 & 0.033 & -4.220 & 0.058 & 3.773 & 0.051 & 12.798 & 1.278 & 0.99\\
  31 & 114.658 & -10.710 & 0.388 & 0.048 & -4.050 & 0.076 & 3.783 & 0.054 & 12.658 & 0.557 & 0.99\\
  32 & 114.644 & -10.706 & 0.387 & 0.034 & -4.102 & 0.054 & 3.684 & 0.047 & 12.516 & 1.294 & 1.0\\
  33 & 114.649 & -10.710 & 0.360 & 0.028 & -4.281 & 0.044 & 3.687 & 0.038 & 14.436 & 0.582 & 0.99\\
  34 & 114.634 & -10.713 & 0.373 & 0.031 & -4.209 & 0.050 & 3.737 & 0.042 & 14.902 & 0.641 & 0.99\\
  35 & 114.653 & -10.721 & 0.354 & 0.025 & -4.151 & 0.041 & 3.745 & 0.035 & 14.337 & 0.594 & 1.0\\
  36 & 114.642 & -10.709 & 0.425 & 0.120 & -4.644 & 0.185 & 3.690 & 0.152 & 17.619 & 1.219 & 0.89\\
  37 & 114.650 & -10.727 & 0.374 & 0.031 & -4.193 & 0.051 & 3.680 & 0.044 & 14.966 & 0.572 & 1.0\\
  38 & 114.579 & -10.749 & 0.465 & 0.110 & -3.828 & 0.174 & 3.455 & 0.142 & 17.488 & 1.044 & 0.93\\
  39 & 114.564 & -10.739 & 0.242 & 0.077 & -4.169 & 0.139 & 3.751 & 0.110 & 16.662 & 0.905 & 0.96\\
  40 & 114.557 & -10.750 & 0.347 & 0.048 & -3.966 & 0.079 & 3.638 & 0.062 & 15.957 & 0.856 & 0.99\\
  41 & 114.584 & -10.751 & 0.357 & 0.139 & -4.208 & 0.220 & 3.853 & 0.175 & 17.892 & 1.124 & 0.95\\
  42 & 114.577 & -10.742 & 0.505 & 0.193 & -3.899 & 0.302 & 3.166 & 0.233 & 18.414 & 1.316 & 0.89\\
  43 & 114.584 & -10.739 & 0.391 & 0.105 & -4.094 & 0.165 & 3.719 & 0.132 & 17.332 & 1.006 & 0.98\\
  44 & 114.602 & -10.725 & 0.268 & 0.047 & -4.268 & 0.075 & 3.716 & 0.060 & 15.907 & 0.773 & 0.96\\
  45 & 114.587 & -10.715 & 0.338 & 0.038 & -4.197 & 0.064 & 3.739 & 0.049 & 15.388 & 0.670 & 0.99\\
  46 & 114.600 & -10.731 & 0.364 & 0.041 & -4.301 & 0.067 & 3.781 & 0.054 & 15.613 & 0.714 & 0.96\\
  47 & 114.604 & -10.729 & 0.399 & 0.050 & -4.091 & 0.082 & 3.769 & 0.065 & 16.075 & 0.781 & 0.99\\
  48 & 114.589 & -10.733 & 0.335 & 0.061 & -4.447 & 0.097 & 3.638 & 0.076 & 16.336 & 0.968 & 0.94\\
  49 & 114.573 & -10.731 & 0.747 & 0.153 & -3.886 & 0.243 & 3.766 & 0.185 & 18.013 & 1.145 & 0.91\\
  50 & 114.562 & -10.731 & 0.420 & 0.078 & -3.996 & 0.134 & 3.769 & 0.099 & 16.919 & 0.972 & 0.98\\
  51 & 114.545 & -10.719 & 0.331 & 0.041 & -4.228 & 0.072 & 3.685 & 0.058 & 15.478 & 0.686 & 0.99\\
  52 & 114.561 & -10.701 & 0.282 & 0.109 & -4.387 & 0.184 & 3.533 & 0.145 & 17.494 & 1.052 & 0.96\\
  53 & 114.560 & -10.709 & 0.244 & 0.154 & -4.282 & 0.248 & 3.643 & 0.188 & 18.065 & 1.178 & 0.95\\
  54 & 114.568 & -10.709 & 0.367 & 0.034 & -4.102 & 0.056 & 3.498 & 0.043 & 15.114 & 0.726 & 0.99\\
  55 & 114.571 & -10.705 & 0.497 & 0.177 & -4.502 & 0.282 & 3.616 & 0.216 & 18.285 & 1.217 & 0.93\\
  56 & 114.551 & -10.708 & 0.111 & 0.172 & -4.076 & 0.276 & 3.466 & 0.208 & 18.250 & 1.429 & 0.95\\
  57 & 114.570 & -10.706 & 0.218 & 0.107 & -4.386 & 0.174 & 3.958 & 0.131 & 17.431 & 1.035 & 0.87\\
  58 & 114.618 & -10.701 & 0.474 & 0.102 & -4.082 & 0.170 & 3.755 & 0.142 & 17.400 & 1.168 & 0.97\\
  59 & 114.610 & -10.719 & 0.435 & 0.153 & -3.889 & 0.235 & 3.708 & 0.180 & 17.955 & 1.118 & 0.96\\
  60 & 114.609 & -10.717 & 0.426 & 0.056 & -4.279 & 0.099 & 3.599 & 0.075 & 16.195 & 0.813 & 0.98\\
\end{tabular}
\end{table*}
\end{center}

 \setcounter{table}{1}

\begin{table*} \caption{continued}
\begin{tabular}{cccccccccccc}
\hline
ID & RA    & DE    & Plx   & e\_{Plx} & pmRA     & e\_{pmRA} & pmDE     & e\_{pmDE} & Gmag  & BP-RP & MP \\
   & (deg) & (deg) & (mas) & (mas)   & (mas/yr) & (mas/yr) & (mas/yr) & (mas/yr) & (mag) & (mag) &    \\
\hline 
  61 & 114.623 & -10.709 & 0.387 & 0.130 & -4.419 & 0.202 & 3.881 & 0.159 & 17.719 & 1.066 & 0.91\\
  62 & 114.606 & -10.713 & 0.355 & 0.027 & -4.099 & 0.046 & 3.633 & 0.037 & 12.972 & 1.234 & 1.0\\
  63 & 114.618 & -10.720 & 0.363 & 0.020 & -3.956 & 0.033 & 3.719 & 0.026 & 13.337 & 0.631 & 0.99\\
  64 & 114.599 & -10.707 & 0.401 & 0.031 & -4.138 & 0.050 & 3.685 & 0.037 & 13.896 & 0.458 & 1.0\\
  65 & 114.605 & -10.710 & 0.427 & 0.026 & -4.094 & 0.045 & 3.736 & 0.036 & 14.253 & 0.487 & 0.99\\
  66 & 114.608 & -10.710 & 0.348 & 0.032 & -4.032 & 0.052 & 3.601 & 0.042 & 15.051 & 0.595 & 0.99\\
  67 & 114.617 & -10.716 & 0.350 & 0.047 & -4.232 & 0.076 & 3.537 & 0.061 & 16.008 & 0.772 & 0.98\\
  68 & 114.621 & -10.717 & 0.390 & 0.029 & -4.031 & 0.046 & 3.607 & 0.037 & 14.793 & 0.622 & 0.99\\
  69 & 114.613 & -10.712 & 0.599 & 0.080 & -4.406 & 0.132 & 4.276 & 0.105 & 15.969 & 0.923 & 0.78\\
  70 & 114.609 & -10.708 & 0.347 & 0.031 & -4.119 & 0.048 & 3.665 & 0.038 & 12.950 & 1.237 & 1.0\\
  71 & 114.608 & -10.698 & 0.382 & 0.029 & -4.061 & 0.049 & 3.586 & 0.039 & 14.754 & 0.534 & 0.99\\
  72 & 114.617 & -10.712 & 0.482 & 0.028 & -4.164 & 0.046 & 3.655 & 0.037 & 14.788 & 0.636 & 0.99\\
  73 & 114.621 & -10.712 & 0.387 & 0.144 & -4.066 & 0.231 & 3.530 & 0.180 & 18.001 & 1.271 & 0.97\\
  74 & 114.598 & -10.708 & 0.432 & 0.135 & -4.084 & 0.205 & 3.731 & 0.160 & 17.584 & 1.021 & 0.97\\
  75 & 114.613 & -10.698 & 0.410 & 0.047 & -3.944 & 0.072 & 3.790 & 0.052 & 15.434 & 0.679 & 0.98\\
  76 & 114.621 & -10.703 & 0.419 & 0.021 & -4.110 & 0.034 & 3.714 & 0.027 & 13.576 & 0.593 & 1.0\\
  77 & 114.615 & -10.709 & 0.372 & 0.020 & -4.143 & 0.032 & 3.732 & 0.026 & 13.354 & 0.480 & 1.0\\
  78 & 114.628 & -10.714 & 0.428 & 0.051 & -4.062 & 0.082 & 3.629 & 0.066 & 16.143 & 0.813 & 0.99\\
  79 & 114.606 & -10.700 & 0.347 & 0.030 & -4.116 & 0.047 & 3.619 & 0.035 & 13.415 & 0.280 & 0.99\\
  80 & 114.649 & -10.695 & 0.349 & 0.026 & -4.138 & 0.052 & 3.642 & 0.035 & 13.618 & 0.452 & 0.99\\
  81 & 114.622 & -10.683 & 0.544 & 0.144 & -4.169 & 0.221 & 3.721 & 0.173 & 17.929 & 1.095 & 0.95\\
  82 & 114.617 & -10.689 & 0.399 & 0.036 & -4.012 & 0.058 & 3.921 & 0.046 & 15.288 & 0.662 & 0.93\\
  83 & 114.615 & -10.691 & 0.281 & 0.037 & -4.053 & 0.061 & 3.679 & 0.049 & 15.443 & 0.685 & 0.98\\
  84 & 114.624 & -10.705 & 0.369 & 0.023 & -4.022 & 0.037 & 3.724 & 0.031 & 14.014 & 0.501 & 1.0\\
  85 & 114.624 & -10.695 & 0.340 & 0.058 & -4.214 & 0.093 & 3.498 & 0.068 & 16.182 & 0.805 & 0.98\\
  86 & 114.629 & -10.703 & 0.354 & 0.023 & -4.099 & 0.036 & 3.620 & 0.029 & 13.879 & 0.505 & 1.0\\
  87 & 114.626 & -10.703 & 0.328 & 0.163 & -4.224 & 0.254 & 3.991 & 0.198 & 18.206 & 1.425 & 0.92\\
  88 & 114.626 & -10.692 & 0.274 & 0.034 & -4.409 & 0.054 & 3.886 & 0.040 & 14.865 & 0.682 & 0.87\\
  89 & 114.648 & -10.693 & 0.302 & 0.032 & -4.234 & 0.059 & 3.729 & 0.044 & 12.959 & 1.261 & 0.98\\
  90 & 114.626 & -10.683 & 0.367 & 0.022 & -4.132 & 0.037 & 3.652 & 0.031 & 13.833 & 0.525 & 1.0\\
  91 & 114.646 & -10.697 & 0.388 & 0.096 & -4.394 & 0.162 & 3.857 & 0.142 & 17.291 & 0.963 & 0.91\\
  92 & 114.636 & -10.685 & 0.518 & 0.114 & -3.497 & 0.170 & 3.580 & 0.131 & 17.198 & 1.180 & 0.79\\
  93 & 114.641 & -10.688 & 0.394 & 0.030 & -4.252 & 0.051 & 3.644 & 0.045 & 14.873 & 0.587 & 0.99\\
  94 & 114.624 & -10.696 & 0.438 & 0.036 & -4.195 & 0.057 & 3.728 & 0.045 & 15.303 & 0.618 & 0.99\\
  95 & 114.629 & -10.695 & 0.309 & 0.022 & -4.317 & 0.036 & 3.676 & 0.029 & 13.958 & 0.590 & 0.97\\
  96 & 114.624 & -10.687 & 0.373 & 0.021 & -4.075 & 0.034 & 3.729 & 0.027 & 13.589 & 0.603 & 1.0\\
  97 & 114.617 & -10.696 & 0.395 & 0.067 & -4.179 & 0.106 & 3.673 & 0.077 & 16.400 & 0.843 & 0.99\\
  98 & 114.616 & -10.697 & 0.323 & 0.038 & -4.170 & 0.059 & 3.738 & 0.043 & 15.033 & 0.573 & 0.99\\
  99 & 114.619 & -10.689 & 0.305 & 0.075 & -4.362 & 0.116 & 4.003 & 0.092 & 16.784 & 0.953 & 0.86\\
  100 & 114.623 & -10.698 & 0.324 & 0.043 & -3.918 & 0.070 & 3.637 & 0.056 & 15.839 & 0.851 & 0.98\\
  101 & 114.627 & -10.698 & 0.372 & 0.026 & -4.145 & 0.041 & 3.657 & 0.032 & 14.324 & 0.464 & 1.0\\
  102 & 114.638 & -10.686 & 0.346 & 0.029 & -4.081 & 0.044 & 3.674 & 0.034 & 13.786 & 0.479 & 1.0\\
  103 & 114.591 & -10.704 & 0.409 & 0.029 & -4.276 & 0.049 & 3.749 & 0.038 & 14.672 & 0.591 & 0.98\\
  104 & 114.583 & -10.679 & 0.472 & 0.154 & -3.765 & 0.259 & 3.732 & 0.202 & 18.009 & 1.328 & 0.95\\
  105 & 114.578 & -10.688 & 0.374 & 0.050 & -4.259 & 0.080 & 3.788 & 0.062 & 16.055 & 0.797 & 0.97\\
  106 & 114.589 & -10.681 & 0.359 & 0.073 & -4.255 & 0.116 & 3.491 & 0.093 & 16.709 & 1.018 & 0.98\\
  107 & 114.592 & -10.682 & 0.417 & 0.063 & -4.038 & 0.100 & 3.789 & 0.080 & 16.549 & 0.982 & 0.98\\
  108 & 114.604 & -10.698 & 0.396 & 0.048 & -4.152 & 0.078 & 3.717 & 0.063 & 16.068 & 0.780 & 0.99\\
  109 & 114.590 & -10.687 & 0.586 & 0.125 & -4.193 & 0.208 & 3.615 & 0.155 & 17.757 & 1.049 & 0.96\\
  110 & 114.596 & -10.691 & 0.412 & 0.038 & -4.396 & 0.062 & 3.982 & 0.049 & 15.575 & 0.762 & 0.87\\
  111 & 114.594 & -10.694 & 0.277 & 0.051 & -4.244 & 0.086 & 3.571 & 0.067 & 16.167 & 0.821 & 0.96\\
  112 & 114.597 & -10.690 & 0.306 & 0.049 & -4.161 & 0.078 & 3.888 & 0.063 & 16.035 & 0.781 & 0.94\\
  113 & 114.601 & -10.690 & 0.654 & 0.198 & -4.219 & 0.298 & 3.644 & 0.245 & 18.395 & 1.283 & 0.94\\
  114 & 114.610 & -10.665 & 0.347 & 0.112 & -4.495 & 0.183 & 3.563 & 0.141 & 17.625 & 1.124 & 0.94\\
  115 & 114.606 & -10.685 & 0.297 & 0.048 & -4.423 & 0.078 & 3.589 & 0.062 & 16.062 & 0.791 & 0.95\\
  116 & 114.612 & -10.684 & 0.339 & 0.022 & -4.103 & 0.036 & 3.672 & 0.029 & 13.886 & 0.506 & 1.0\\
  117 & 114.602 & -10.676 & 0.312 & 0.057 & -4.110 & 0.102 & 3.660 & 0.089 & 16.293 & 0.834 & 0.99\\
  118 & 114.599 & -10.680 & 0.434 & 0.028 & -4.181 & 0.047 & 3.656 & 0.037 & 14.857 & 0.554 & 0.99\\
  119 & 114.620 & -10.676 & 0.371 & 0.078 & -4.266 & 0.127 & 3.793 & 0.100 & 16.993 & 0.940 & 0.96\\
  120 & 114.611 & -10.674 & 0.363 & 0.020 & -4.160 & 0.032 & 3.709 & 0.026 & 13.275 & 1.253 & 1.0\\
 \end{tabular}
 \end{table*}

  \setcounter{table}{1}

\begin{table*} \caption{continued}
\begin{tabular}{cccccccccccc}
\hline
ID & RA    & DE    & Plx   & e\_{Plx} & pmRA     & e\_{pmRA} & pmDE     & e\_{pmDE} & Gmag  & BP-RP & MP \\
   & (deg) & (deg) & (mas) & (mas)   & (mas/yr) & (mas/yr) & (mas/yr) & (mas/yr) & (mag) & (mag) &    \\
\hline 
  121 & 114.602 & -10.671 & 0.341 & 0.035 & -4.148 & 0.060 & 3.910 & 0.050 & 15.258 & 0.709 & 0.93\\
  122 & 114.590 & -10.674 & 0.400 & 0.136 & -4.046 & 0.220 & 3.625 & 0.176 & 17.861 & 1.087 & 0.97\\
  123 & 114.617 & -10.686 & 0.203 & 0.134 & -4.059 & 0.214 & 3.326 & 0.168 & 17.879 & 1.124 & 0.95\\
  124 & 114.588 & -10.674 & 0.340 & 0.023 & -4.153 & 0.041 & 3.653 & 0.032 & 13.705 & 0.502 & 0.99\\
  125 & 114.615 & -10.682 & 0.527 & 0.138 & -4.352 & 0.218 & 3.636 & 0.171 & 17.947 & 1.122 & 0.95\\
  126 & 114.616 & -10.683 & 0.329 & 0.086 & -4.199 & 0.137 & 3.378 & 0.107 & 17.100 & 0.921 & 0.97\\
  127 & 114.608 & -10.690 & 0.339 & 0.028 & -4.054 & 0.045 & 3.771 & 0.036 & 14.704 & 0.541 & 0.99\\
  128 & 114.602 & -10.685 & 0.301 & 0.089 & -4.242 & 0.139 & 3.961 & 0.118 & 17.121 & 0.952 & 0.91\\
  129 & 114.682 & -10.707 & 0.417 & 0.069 & -4.223 & 0.111 & 3.794 & 0.092 & 16.648 & 0.891 & 0.97\\
  130 & 114.671 & -10.689 & 0.369 & 0.060 & -4.055 & 0.097 & 3.703 & 0.081 & 16.359 & 0.859 & 0.99\\
  131 & 114.671 & -10.703 & 0.352 & 0.024 & -4.244 & 0.044 & 3.599 & 0.038 & 14.072 & 0.480 & 0.99\\
  132 & 114.672 & -10.708 & 0.236 & 0.138 & -4.218 & 0.237 & 3.652 & 0.202 & 17.900 & 1.215 & 0.96\\
  133 & 114.690 & -10.696 & 0.402 & 0.074 & -4.612 & 0.117 & 3.625 & 0.099 & 16.737 & 0.891 & 0.88\\
  134 & 114.682 & -10.692 & 0.494 & 0.064 & -4.255 & 0.102 & 3.820 & 0.085 & 16.476 & 0.861 & 0.93\\
  135 & 114.675 & -10.686 & 0.323 & 0.105 & -4.230 & 0.164 & 3.573 & 0.135 & 17.394 & 1.035 & 0.97\\
  136 & 114.681 & -10.670 & 0.165 & 0.147 & -4.486 & 0.232 & 3.834 & 0.190 & 17.967 & 1.142 & 0.9\\
  137 & 114.696 & -10.674 & 0.336 & 0.026 & -4.163 & 0.042 & 3.723 & 0.037 & 13.008 & 1.230 & 1.0\\
  138 & 114.687 & -10.666 & 0.283 & 0.127 & -4.205 & 0.199 & 3.914 & 0.162 & 17.716 & 1.070 & 0.93\\
  139 & 114.697 & -10.679 & 0.373 & 0.079 & -4.008 & 0.127 & 3.442 & 0.107 & 16.905 & 0.922 & 0.98\\
  140 & 114.694 & -10.685 & 0.422 & 0.069 & -4.063 & 0.107 & 3.571 & 0.091 & 16.534 & 0.856 & 0.99\\
  141 & 114.647 & -10.681 & 0.328 & 0.030 & -4.135 & 0.050 & 3.653 & 0.040 & 12.994 & 0.605 & 0.99\\
  142 & 114.638 & -10.678 & 0.364 & 0.025 & -4.086 & 0.040 & 3.663 & 0.034 & 13.961 & 0.551 & 1.0\\
  143 & 114.636 & -10.676 & 0.190 & 0.098 & -4.302 & 0.157 & 3.516 & 0.132 & 17.317 & 0.995 & 0.95\\
  144 & 114.636 & -10.674 & 0.263 & 0.095 & -4.151 & 0.149 & 3.612 & 0.128 & 17.093 & 1.126 & 0.97\\
  145 & 114.644 & -10.677 & 0.390 & 0.090 & -4.452 & 0.159 & 3.384 & 0.135 & 17.185 & 1.059 & 0.95\\
  146 & 114.686 & -10.654 & 0.404 & 0.031 & -4.174 & 0.050 & 3.628 & 0.043 & 14.391 & 0.489 & 0.99\\
  147 & 114.637 & -10.668 & 0.345 & 0.021 & -4.153 & 0.038 & 3.629 & 0.033 & 13.096 & 1.263 & 0.99\\
  148 & 114.631 & -10.673 & 0.356 & 0.021 & -4.049 & 0.034 & 3.698 & 0.029 & 13.127 & 1.275 & 1.0\\
  149 & 114.643 & -10.664 & 0.428 & 0.064 & -3.943 & 0.104 & 3.778 & 0.089 & 16.527 & 0.889 & 0.98\\
  150 & 114.626 & -10.641 & 0.427 & 0.033 & -4.300 & 0.054 & 3.683 & 0.045 & 15.021 & 0.598 & 0.97\\
  151 & 114.626 & -10.655 & 0.403 & 0.044 & -4.237 & 0.069 & 3.594 & 0.059 & 15.650 & 0.728 & 0.99\\
  152 & 114.642 & -10.639 & 0.248 & 0.191 & -3.819 & 0.287 & 3.494 & 0.227 & 18.256 & 1.097 & 0.94\\
  153 & 114.657 & -10.632 & 0.355 & 0.027 & -4.137 & 0.046 & 3.675 & 0.038 & 14.513 & 0.492 & 1.0\\
  154 & 114.663 & -10.636 & 0.536 & 0.072 & -4.255 & 0.114 & 3.746 & 0.099 & 16.676 & 0.871 & 0.94\\
  155 & 114.659 & -10.648 & 0.572 & 0.109 & -4.110 & 0.172 & 3.561 & 0.143 & 17.477 & 1.033 & 0.96\\
  156 & 114.544 & -10.691 & 0.306 & 0.054 & -4.188 & 0.092 & 3.582 & 0.070 & 16.197 & 0.834 & 0.98\\
  157 & 114.546 & -10.695 & 0.538 & 0.158 & -3.852 & 0.258 & 4.130 & 0.190 & 18.085 & 1.166 & 0.85\\
  158 & 114.579 & -10.669 & 0.327 & 0.185 & -4.115 & 0.294 & 3.540 & 0.226 & 18.300 & 1.407 & 0.96\\
  159 & 114.566 & -10.678 & 0.486 & 0.036 & -4.135 & 0.060 & 3.657 & 0.048 & 15.451 & 0.749 & 0.99\\
  160 & 114.571 & -10.680 & 0.492 & 0.045 & -4.242 & 0.074 & 3.899 & 0.057 & 15.762 & 0.926 & 0.88\\
  161 & 114.568 & -10.675 & 0.410 & 0.074 & -4.212 & 0.130 & 3.543 & 0.101 & 16.810 & 0.916 & 0.98\\
  162 & 114.571 & -10.688 & 0.495 & 0.199 & -4.306 & 0.327 & 3.228 & 0.248 & 18.469 & 1.277 & 0.94\\
  163 & 114.558 & -10.674 & 0.371 & 0.055 & -4.130 & 0.091 & 3.751 & 0.071 & 16.239 & 0.835 & 0.99\\
  164 & 114.591 & -10.648 & 0.420 & 0.139 & -4.066 & 0.216 & 3.781 & 0.173 & 17.913 & 1.218 & 0.96\\
  165 & 114.599 & -10.636 & 0.429 & 0.028 & -4.054 & 0.047 & 3.680 & 0.041 & 14.683 & 0.588 & 0.99\\
  166 & 114.596 & -10.643 & 0.315 & 0.063 & -4.087 & 0.102 & 3.672 & 0.080 & 16.529 & 0.967 & 0.99\\
  167 & 114.623 & -10.633 & 0.370 & 0.032 & -4.066 & 0.054 & 3.774 & 0.047 & 14.859 & 0.596 & 0.99\\
  168 & 114.630 & -10.624 & 0.363 & 0.036 & -4.216 & 0.065 & 3.580 & 0.055 & 15.420 & 0.664 & 0.99\\
 \end{tabular}
 \end{table*} 


\section{Conclusions}

We present here a new analysis study for the open star cluster Melotte 72 depending on the Gaia DR2 database, estimating the most photometric parameters of the cluster under study. Melotte 72 is defined as a compressed small cluster in the constellation Monoceros, which lies approximately 1.3 degrees south-west of $\alpha$ Mon. The membership probability was determined using ASteCA (the Automated Stellar Cluster Analysis) code. Our work compared with previous studies by Hasegawa et al. (2008), Piatti et al. (2010), and Cantat-Gaudin et al. (2020). We found that our work agrees well with Cantat-Gaudin et al. (2020), as shown in Fig. 10 and listed in Table 2. For verifying our results on the ground-based observations, we got the Piatti et al. (2010)’s members that matched with our members’ coordinates on (V$\sim$V-I) diagram. The fit agrees well with ours in age, distance modulus, and reddening. All results are listed in Table 1, where the current and previous results can be compared.

{\bf Acknowledgments} 

This work is a part of a project called "IMHOTEP" program No. 42088ZK between Egypt and France. We are grateful to the Academy of Scientific Research and to the Service of scientific cooperation in both countries for giving us the opportunity to work through this project and provide the necessary support to accomplish our scientific goals. This work has made use of data from the European Space Agency (ESA) mission Gaia processed by the Gaia Data Processing and Analysis Consortium (DPAC), \\ (www.cosmos.esa.int/web/gaia/dpac/consortium).\\ Funding for the DPAC has been provided by national institutions, in particular, the institutions participating in the Gaia Multilateral Agreement.






\begin{thebibliography}{}

\bibitem{0001} Caffau, E., Maiorca, E., Bonifacio, P., et al.: 2009, A\&A, 498, 877
\bibitem{0001} Caffau, E., Ludwig, H., Steffen, M., et al.: 2011, Solar Physics, 268 (2), 255
\bibitem{0001} Cantat-Gaudin, A., et al.: 2018, A\&A, 618, A93
\bibitem{0001} Cantat-Gaudin, T., et al. 2020, A\&A, 640, A1
\bibitem{0001} Castro-Girard, A., et al.: 2020, A\&A, 635, A45
\bibitem{0001} Collinder P.: 1931, Medd. Lunds Astron. Obs. No. 2
\bibitem{0001} Dias, W., Alessi, B., Moitinho, A., Lepine, J.R.: 2002, A\&A 389, 871
\bibitem{0001} Dias, W., Monteiro H., Caetano, T., et al.: 2014, A\&A, 564-A, 79
\bibitem{0001} Elmegreen B.G., 1999, ApJ, 515, 323
\bibitem{0001} Elmegreen B.G., 2000, ApJ, 539, 342
\bibitem{0001} Gaia Collaboration et al.: 2018, A\&A, 616, A1
\bibitem{0001} Gilmore, G., Randich, S., Asplund, M., et al.: 2012, The Messenger, 147, 25
\bibitem{0001} Hasegawa T., Sakamoto T., Malasan H. 2008, Publ. Astron. Soc. Japan 60, 1267
\bibitem{0001} Hendy, Y. H. M.: 2018, NRIAG Journal of Astronomy and Geophysics, 7, Issue 2, 180
\bibitem{je01} Jeffries, R., Thurston, M., Hambly, N.: 2001, A\&A, 375, 863
\bibitem{0001} Kharchenko, N.V., Piskunov, A.E., Röser, S., et al.: 2004, Astron. Nachr., 325, 740
\bibitem{0001} Kharchenko, N. V., Piskunov, A. E., Schilbach, E., Röser S., Scholz, R.: 2013, A\&A, 558, A53
\bibitem{ki66} King, I.: 1966, AJ, 71, 64
\bibitem{ki66} Kroupa, Pavel, 2001, MNRAS, 322, 231K
\bibitem{0001} Lynga, G.: 1982, A\&A, 109, 213
\bibitem{0001} Marigo, et al.: 2017, ApJ, 835, 77
\bibitem{0001} Moraux, E.: 2016, EAS Publications Series, 80, 7311
\bibitem{0001} Netopil, M., Paunzen, E., Carraro, G.: 2015, A\&A, 582, A19
\bibitem{0001} Phelps R. L., and Janes K.A., 1993, AJ, 106, 1870
\bibitem{0001} Perren G. I., Vazquez R. A., and Piatti A. E., 2015, A\&A 576 
\bibitem{0001} Perren G. I., Giorgi E. E., Moitinho A., et al. 2020, A\&A 637, A95
\bibitem{0001} Peterson, C.\& King, I.: 1975, AJ, 80, 427
\bibitem{0001} Piatti, A.E., et al.: 2010, MNRAS, 402, 2270
\bibitem{0001} Piskunov A. E., Belikov A.N., Kharchenko N. V., et al., 2004, MNRAS, 349, 1449
\bibitem{0001} Richtler Tom, 1994, A\&A, 287, 517R
\bibitem{0001} Roeser, S., Demleitner, M., Schilbach, E.: 2010, AJ, 139, 2440
\bibitem{sa55} Salpeter, E.: 1955, ApJ, 121, 161
\bibitem{0001} Sampedro, L., Dias, W., Alfaro, E., Monteiro, H., Molino, A.: 2017, MNRAS, 470, 3937
\bibitem{0001} Scalo J.M., 1986, Fund. Cosmic Phys, 11, 1
\bibitem{sh06} Sharma, S., Pandey A., Ogura, K., et al.: 2006, AJ, 132, 1669
\bibitem{0001} Spitzer, Lyman Jr., Hart Michael, H.: 1971, AJP, 164, 399 
\bibitem{0001} Steffi, X.,Yen Reffert, S., Röser, S., et al.: 2018, IAU Symposium, 330, 281
\bibitem{ta11} Tadross, A. L.: 2011, JKAS, 44, 1
\bibitem{0001} Tadross, A.L.: 2018, RAA, 18, 158
\bibitem{0001} Yadav R.K.S., and Sagar R., 2002, MNRAS, 337, 133
\bibitem{0001} Yadav R.K.S., and Sagar R., 2004a, MNRAS, 349, 1481
\end{thebibliography}
\end{document}